# NOVEL DEEP LEARNING FRAMEWORK FOR BOVINE IRIS SEGMENTATION

*Heemoon Yoon[a], Mira Park[a], Sang-Hee Lee[b]*

[a]School of ICT, University of Tasmania, Hobart, Australia
[b]College of Animal Life Sciences, Kangwon National University, Chuncheon, South Korea

## ABSTRACT

Iris segmentation is the initial step to identify biometric of animals to establish a traceability system of livestock. In this study, we propose a novel deep learning framework for pixel-wise segmentation with minimum use of annotation labels using BovineAAEyes80 public dataset. In the experiment, U-Net with VGG16 backbone was selected as the best combination of encoder and decoder model, demonstrating a 99.50% accuracy and a 98.35% Dice coefficient score. Remarkably, the selected model accurately segmented corrupted images even without proper annotation data. This study contributes to the advancement of the iris segmentation and the development of a reliable DNNs training framework.

## 1. INTRODUCTION

Accurate identification of animal entities applies to the whole process of livestock food production, so it is very important for establishing a traceability system of the food supply chain from farm to our table. [1, 2]. Reliable animal identification methodologies allow monitoring each stage of production while minimizing trade losses and ensuring animal ownership at the same time. In order to implement such a tracking system, a robust identification methodology is required [3] because the damage caused by the failure of the tracking system is enormous. The damage is linked to food safety, which can put customers at risk and cause serious economic problems [4].

To mitigate these potential hazards, there have been several traditional methods to reserve identity or animals: Ear notching, tattoos, tags and branding would be some of the permanent methods used in traditional ways, but these are easy to be thieved, frauded, and duplicated [5]. As an alternate of the traditional methods, Radio Frequency Identification (RFID) tag has been implemented [6]. RFID allows animal to be registered in computer systems and can be identified by scanning RFID tag electrically. However, this tag hold limitation that it is invasive approach and can also be frauded by getting manipulated in the inside of the system [7]. For the most recent approach to resolve the issue, biometrics such as retinal vascular patterns (RVPs) [8], muzzle [9, 10], iris [11, 12] have been proposed. These methods utilizing biometrics are regarded as highly reliable in the identification of an entity as it is one of the most accurate and stable biometric modalities during the lifetime of animals [3, 13].

With the advent of deep neural networks (DNNs) [14], there have been several attempts to identify anatomical parts of an animal using deep learning technologies [10, 15-17]. Among the deep learning technologies, the segmentation technique allows us to classify objects within a given image in a pixel-wise manner. As segmentation of iris from the whole eye image is an essential step to initiate iris identification, using an elaborate and accurate segmentation technique is key to success iris recognition [16].

In this study, we elaborate on bovine iris segmentation using our novel framework. Our framework will develop multiple segmentation models by training publicly available bovine iris datasets, BovineAAEyes80 [18], and comparing combinations of the state-of-art deep learning techniques. Since iris datasets are rare and limited in their format like other biometric datasets, we propose a framework enabling us to develop models using minimum input datasets: region of interests (ROIs) labels and RGB images. This study will not only contribute to the advancement of iris identification using DNNs but also to the development of a reliable DNNs training framework assisting decision making of the most suitable combination of DNNs models for biometric images.

## 2. METHODOLOGY

### 2.1. Framework Overview

The proposed framework starts with data collection. The input data must contain pairs of image and annotation data (Figure 1a). After collecting pairs of data, the data preparation stage comprises of 2 steps: data splitting and augmentation selection. Data must be split into training, validation, and test dataset while desirably keeping mutual exclusion so that each of the training, validation, and testing stages can be conducted with unseen dataset respectively. The data splitting step needs careful consideration to be equally distributed in terms of both quality and quantity since this step can affect the result of the trained model [19]. Augmentation selection step can be varied to the traits of the dataset (Figure 1b). After selecting the augmentation options, training of 15 combination of DNNs models is conducted using 3 different encoder backbones (VGG16 [20], ResNet50 [21], MobileNet [22]) and 5 different segmentation decoder DNNs (FCN8, FCN16, FCN32 [23], U-Net [24], SegNet [25]). As a result, trained 15 models are returned and assessed. In this study, since we performed 5-fold repeat experiments to ensure the reliability of the training result (Table 1), 75 models with 15 combinations of models were returned and. Evaluation step is subjected to the 75 models achieved from the previous steps with an unseen test dataset (Figure 1c). This framework will return detailed information about each of the models including training and test results with various metrics such as accuracy, precision, recall, IOU and Dice-coefficient [31], and graphical segmentation result (Figure 1d).

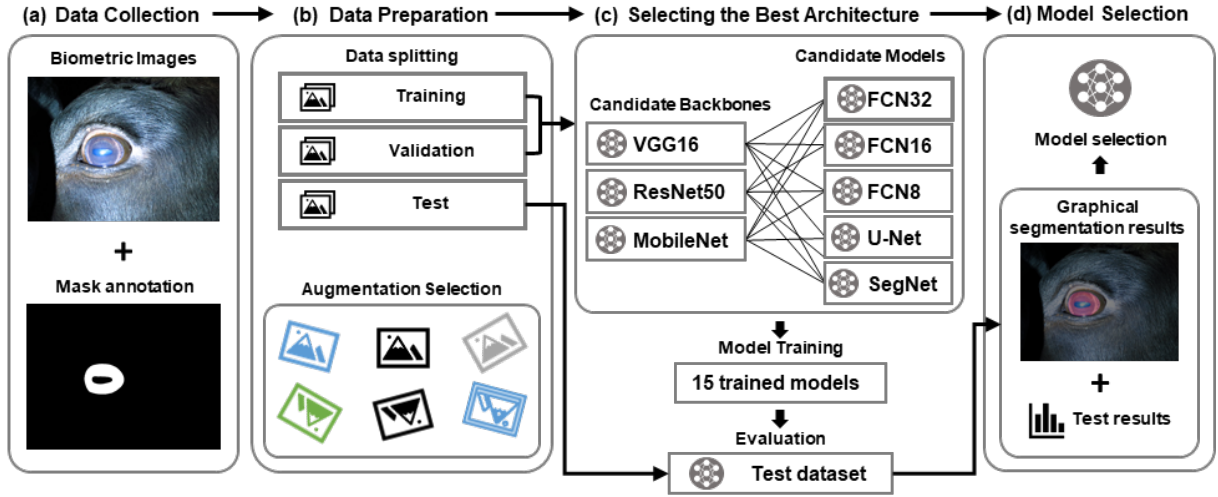

**Fig. 1:** Scheme of model training for selection of the best combination of segmentation models for biometric images.

**Table 1:** Distribution of the dataset for 5-fold cross validation

| Dataset | Fold | Images | Eye ID |
|---|---|---|---|
| Train | Fold 1 | 12 | 3, 7 |
|  | Fold 2 | 13 | 4 |
|  | Fold 3 | 13 | 5, 6 |
|  | Fold 4 | 12 | 8, 9 |
|  | Fold 5 | 22 | 10, 11 |
| Test | - | 8 | 1, 2 |
| Total |  | 80 images | 11 eyes |

**2.2. Model Training Environment and Configuration**

With reference to previous literatures, we compared five candidates, FCN32, FCN16, FCN8, U-Net, and SegNet, to find the most reliable architecture for anatomical segmentation (Figure 1c). All configurations were set to be equal for a fair comparison, minimizing any possible variants between model training processes. After several attempts, the training hyperparameters were experimentally determined as follows: training for 100 epochs with 128 steps per epoch, a learning rate of 0.001 optimized using an Adam optimizer, and a batch size of 4. These trained models automatically generated anatomical ROIs from input test images. After training and evaluation with statistical performance measures, such as the Dice coefficient and accuracy [26], the statistical result is returned with csv format and analysed within the framework system.

Model training was conducted on Anaconda 4.10.1, running 64bit Ubuntu Linux 20.04.3 LTS and Python v3.8.8. TensorFlow-GPU v2.7.0 and CUDA 11.4 were used to accelerate the DNNs framework's training process on RTX3090 24GB and Keras v2.7.0 was used as a Python deep learning application programming interface (API). In this BovineAAEyes80, brightness ±10 and rotation ± 40° augmentation are applied to cover variations that can be happened in the capturing environment such as non-cooperative behaviour of bovine and changes in lighting condition [18].

**2.3. Model Evaluation**

The classification performance of the trained model was evaluated using the following metrics: accuracy, recall, precision, Intersection over Union (IoU) and Dice coefficient [26]. Compared to the reference annotation, each pixel is counted as one of four possible outcomes: true positive (TP), true negative (TN), false positive (FP) or false negative (FN) from which these metrics are followed as previous study [27].

**3. RESULTS**

Figure 2 shows the learning curves of model training. In the training curve of VGG16 (Figure 2a and 2b), all of the FCN series are not stable during the training time while FCN32 records the highest loss and lowest accuracy. SegNet and U-Net show comparably stable loss decrease and accuracy increase at most of the training time. In the training curve of

**Table 2:** Test result of the trained models with unseen test dataset

| Decoder | Encoder | Dice | IOU | Accuracy | Recall | Precision |
|---|---|---|---|---|---|---|
| FCN8 | VGG16 | 97.90 ± 0.25[ab] | 95.97 ± 0.45[ab] | 99.37 ± 0.07[a] | 96.81 ± 0.44[ab] | 99.06 ± 0.12[ab] |
| | ResNet50 | 98.31 ± 0.44[a] | 96.75 ± 0.81[a] | 99.49 ± 0.13[a] | 97.50 ± 0.76[a] | 99.17 ± 0.09[a] |
| | MobileNet | 97.14 ± 0.16[abc] | 94.57 ± 0.29[abc] | 99.15 ± 0.04[a] | 95.52 ± 0.37[abc] | 98.91 ± 0.17[ab] |
| FCN16 | VGG16 | 96.91 ± 0.41[abc] | 94.17 ± 0.72[abc] | 99.08 ± 0.12[a] | 95.45 ± 0.63[abc] | 98.48 ± 0.28[abc] |
| | ResNet50 | 96.38 ± 1.20[bc] | 93.39 ± 2.02[bc] | 98.96 ± 0.32[a] | 94.46 ± 1.92[abcd] | 98.67 ± 0.22[ab] |
| | MobileNet | 94.44 ± 0.19[de] | 89.93 ± 0.31[de] | 98.39 ± 0.05[a] | 91.90 ± 0.53[de] | 97.40 ± 0.49[c] |
| FCN32 | VGG16 | 89.70 ± 0.21[f] | 81.77 ± 0.25[f] | 93.96 ± 1.07[c] | 88.73 ± 1.20[f] | 91.13 ± 1.30[e] |
| | ResNet50 | 94.00 ± 0.51[e] | 89.23 ± 0.83[e] | 98.29 ± 0.13[ab] | 90.82 ± 0.77[ef] | 97.81 ± 0.36[bc] |
| | MobileNet | 88.51 ± 0.86[f] | 81.11 ± 1.16[f] | 96.96 ± 0.18[b] | 83.61 ± 1.18[g] | 95.43 ± 0.43[d] |
| SegNet | VGG16 | 96.45 ± 0.75[bc] | 93.40 ± 1.29[bc] | 98.97 ± 0.20[a] | 94.04 ± 1.21[bcd] | 99.24 ± 0.17[a] |
| | ResNet50 | 98.04 ± 0.54[ab] | 96.25 ± 1.01[ab] | 98.05 ± 1.06[ab] | 96.85 ± 1.10[ab] | 99.36 ± 0.13[a] |
| | MobileNet | 95.73 ± 0.33[dc] | 92.09 ± 0.56[cd] | 98.77 ± 0.09[a] | 92.83 ± 0.61[cde] | 99.12 ± 0.09[ab] |
| U-Net | VGG16 | 98.34 ± 0.49[a] | 96.81 ± 0.90[a] | 99.47 ± 0.81[a] | 97.38 ± 0.83[a] | 99.37 ± 0.13[a] |
| | ResNet50 | 98.26 ± 0.42[a] | 96.66 ± 0.78[a] | 99.18 ± 0.27[a] | 97.11 ± 0.82[a] | 99.52 ± 0.09[a] |
| | MobileNet | 98.35 ± 0.24[a] | 96.80 ± 0.45[a] | 99.50 ± 0.07[a] | 97.20 ± 0.47[a] | 99.57 ± 0.07[a] |

Dice: Dice coefficient, the values are represented as mean ± standard error mean (P<0.05)

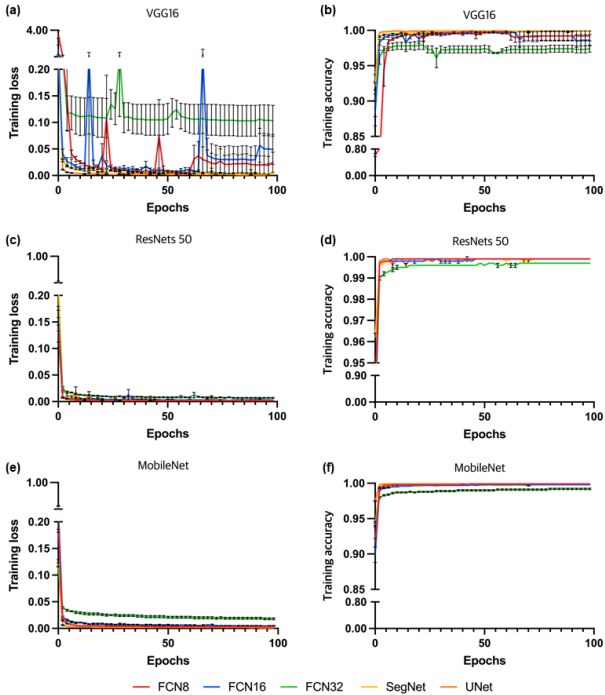

**Fig. 2:** Learning curves of model training.

ResNet50 (Figure 2c and 2d) and MobileNet (Figure 2e and 2f), decent training records for both accuracy and loss are shown with little fluctuation compared with VGG16 while FCN32 still shows the least performance among the models.

Table 2 shows the test result of the trained models with an unseen test dataset. In Table 2, U-Net with MobileNet backbone shows the best score at Dice coefficient (98.35 ± 0.54%), Accuracy (99.50 ± 0.16%) and Precision (99.57 ± 0.16%) while U-Net with VGG16 backbone shows the best IOU score (96.81 ± 2.01%) which is slightly (0.01%) better than the U-Net with MobileNet. In the Table 3, inference time of U-Net with MobileNet demonstrates the fastest mean record (116.3 ± 2.5 ms).

**Table 3**: Inference time comparison of trained models with unseen test dataset.

| Decoder | Encoder | Times (ms) | | |
|---|---|---|---|---|
| | | Mean | Min. | Max |
| FCN8 | VGG16 | 133.3 ± 1.0 | 126.9 | 151.1 |
| | ResNet50 | 180.1 ± 1.6 | 167.8 | 211.8 |
| | MobileNet | 156.8 ± 7.1 | 124.8 | 272.1 |
| FCN16 | VGG16 | 131.6 ± 0.6 | 127.3 | 144.4 |
| | ResNet50 | 182.5 ± 1.3 | 168.2 | 202.7 |
| | MobileNet | 136.7 ± 1.5 | 125.6 | 158.7 |
| FCN32 | VGG16 | 135.1 ± 1.5 | 127.5 | 165 |
| | ResNet50 | 184.5 ± 1.5 | 169.2 | 207.5 |
| | MobileNet | 132.8 ± 2.8 | 125.6 | 158.2 |
| SegNet | VGG16 | 129.2 ± 3.5 | 107.9 | 182.3 |
| | ResNet50 | 126.5 ± 2.1 | 113.2 | 168.6 |
| | MobileNet | 122.8 ± 3.2 | 102.8 | 157.8 |
| U-Net | VGG16 | 125.6 ± 2.1 | 108.2 | 177.8 |
| | ResNet50 | 143.3 ± 7.8 | 113.4 | 347.5 |
| | MobileNet | **116.3 ± 2.5** | 100.8 | 152.8 |

Dice: Dice coefficient, the values are represented as mean ± standard error mean.

According to the records, U-Net with MobileNet can be regarded as the most suitable model for the dataset. However, as it can be seen from Figure 3, there are significant differences in the size of a segmentation unit of pixels between backbones due to the extracted feature map size of each encoder architecture. This should be considered at the model selection along with the numeric scores because the application of the DNNs model can differ for each domain. In the iris segmentation, since the fine segmentation result of the edge of the iris boundaries is emphasized, the second best scored model, U-Net with VGG16, is selected as the best model due to its dense boundary segmentation. The model selection is based on median values of the Dice coefficient in the results of 5-fold cross validations

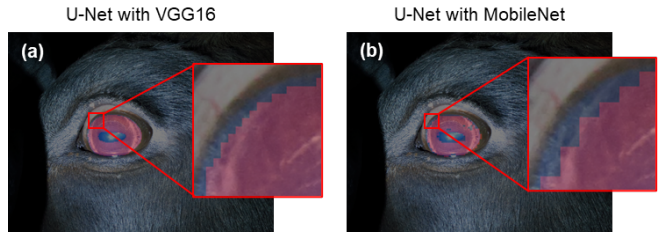

**Fig. 3**: Comparison of the size of a segmentation unit of pixels

## 4. DISCUSSION

The proposed framework suggests a comprehensive solution for the analysis of multiple DNNs model combinations to segment animal iris with robust comparison information. However, there are several potential areas of improvement needed to be applied for further application.

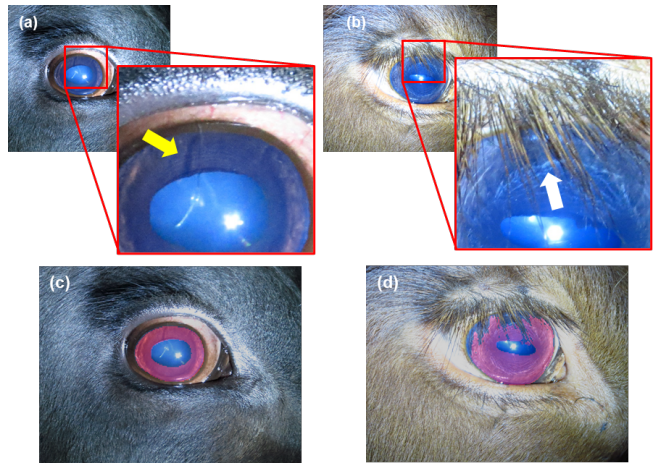

**Fig. 4:** Common corruptions in iris image.

In Figure 4, common corruptions in iris image are described. Minor corruption which distorts iris image can be caused by many reasons such as dust in the spots, stains over the lens, eyelash or fur of the animal in the eye (Figure 4a) and unwanted light spots [16]. In the result, this minor corruption was not reflected in the segmentation result (Figure 4c). However, this issue is needed to be resolved in the further step to mitigate the false information within the iris image. On the contrary, major corruption is generally caused by a relatively large part of their body such as occlusion of eyelashes and eyelids. As it is mentioned in other studies, this major corruption can impede accurate identification [12, 18, 28]. However, the selected best model accurately segmented the corrupted image by excluding the occlusion. This could have not been calculated correctly in the result because the annotation labels used in the model training do not provide that much pixel-wise accurate segmentation ground truth. This seems remarkable compared with other studies using image processing techniques because it segmented the exact iris area exclusively without any preprocessing or postprocessing technique with the model's standalone knowledge even though the corruption was not annotated in the given labels.

## 5. CONCLUSION

With the proposed framework, iris segmentation for identification of animal biometrics was performed utilising the standalone knowledge in the trained DNNs along with robust comparison information provided to assist model determination for the given dataset. As a result, the model selected as the best combination of encoder and decoder model, U-Net with VGG16 backbone, demonstrated a 99.50% accuracy and a 98.35% Dice coefficient score on the unseen test dataset.

This study contributes to the initial step of iris identification to improve the animal tracking system by suggesting a novel framework of DNNs training for pixel-wise segmentation with minimum use of annotation labels. For the reliable comparison of various combinations of DNNs models to select the most suitable DNNs model combination, this approach offers multiple metrics commonly used in the evaluation of segmentation including visual references, so it is unbiased and consistent selection of model. This has the potential to improve accessibility of DNNs technology for layman of DNNs and accelerate inter-study comparison, as well as reducing the variation in current manual model selection. Following this study, the authors plan to improve the framework in the areas of model selection, image segmentation, machine learning, animal biometrics, and multi-resolution imaging. The goal of future research is to develop techniques and skills that can be applied to animal tracking, image recognition, and artificial intelligence applications.

## 6. DATA AND CODE AVAILABILITY

The code used in this study is publicly available from the following link:
https://github.com/boguss1225/IrisSegmentationKeras

The dataset (BovineAAEyes80) used in this study is publicly available from the following link:
https://github.com/juanilarregui/BovineAAEyes80